\documentclass{article}

\usepackage{spconf}
\usepackage{amsmath}
\usepackage{graphicx}
\usepackage{booktabs}
\usepackage{hyperref}
\usepackage{multirow}

\hypersetup{hidelinks}

\graphicspath{{figures/}}

\title{Averaging Collusion Breaks Traceability in Personalized Neural Audio Watermarking}
\name{Yangming Liu$^{*}$, Haiyun Li$^{*}$, Dongfu Song$^{*}$, Ziang Xu,
Jiehui Luo, Yining Zhang, and Zhiyong Wu$^{\dagger}$
\thanks{$^{*}$Equal contribution. $^{\dagger}$Corresponding author.}}

\address{Tsinghua University, Shenzhen, China}

\begin{document}

\ninept
\maketitle

\begin{abstract}
Personalized neural audio watermarking systems embed user-specific payloads into audio to trace the user associated with a leaked or misused copy. Existing evaluations mainly assume access to a single personalized copy, but real-world distribution may expose multiple copies of the same recording. We find that simple sample-wise averaging breaks traceability: averaging only two copies produces tracing-failure rates of 87.7--99.7\% across the five systems while largely preserving audio quality. Decoder outputs show distinct responses to coalition payload composition across systems, with high majority agreement in TimbreWM but substantially lower agreement in WMCodec. In a more sophisticated payload-aware setting, mixture weighting can further steer the decoder toward a selected nonmember, achieving a target success rate of 92.3\% in TimbreWM. Preliminary confidence screening rejects many collusion-induced assignments while retaining most valid single-copy outputs. Code and results are available at \url{https://github.com/Silwan5z/neural-audio-watermark-collusion-attack}. This work highlights multi-copy traceability as an additional evaluation dimension for personalized neural audio watermarking alongside conventional single-copy robustness.
\end{abstract}


\section{Introduction}

Personalized audio watermarking embeds a user-specific payload in each distributed copy. If a copy is leaked or misused, the rights holder can decode its payload to identify the responsible user and hold that user accountable. A tracing failure occurs when the decoded payload matches no registered user or points to an innocent one, potentially causing substantial financial and reputational harm.

Recent advances in neural audio watermarking have enabled robust and imperceptible embedding of identity payloads through learned encoders and decoders. Existing watermarking systems use different embedding strategies, including waveform- or spectrogram-domain message encoding \cite{sanroman2024audioseal,chen2023wavmark,liu2024timbre} and latent or codec-based watermark integration \cite{li2025voicemark,zhou2025wmcodec}. Current evaluations mainly study distortions or attacks applied to one watermarked copy at a time, including compression, resampling, filtering, noise, watermark removal, and forgery \cite{liu2024audiomarkbench,ozer2025rawbench,li2024ideaw}, as well as desynchronization and replay attacks \cite{li2024draw} and watermark overwriting \cite{yao2026overwriting}. These evaluations characterize robustness to individual signal manipulations but generally assume that the attacker operates on only one copy.

This assumption does not account for scenarios in which an attacker obtains multiple valid personalized copies of the same content. With multiple copies, the attacker does not need to estimate or remove the watermark; instead, the attacker can directly combine legitimate copies to interfere with identity decoding. A natural way to combine these copies is sample-wise averaging: personalized copies share the same audio content and differ mainly in their user-specific watermark signals, allowing averaging without knowledge of the watermark algorithm or model architecture. This operation is easy to perform, but it can alter the decoded identity and compromise traceability, leading to untraceable leakage or misuse.

Collusion has long been considered in classical spread-spectrum watermarking \cite{cox1997spread} and multimedia fingerprinting \cite{trappe2003anticollusion,boneh1998fingerprinting,tardos2003fingerprint,celik2004prewarping,tirkel2011audio}, where fingerprints are generated by predefined codes or watermark signals and analyzed under known collusion models. In contrast, neural watermarking systems embed payloads into audio waveforms and recover them using learned decoders. This leaves an end-to-end question that classical code analysis alone does not answer: when valid neural-watermarked waveforms are averaged, how does the learned decoder map the mixture back to a payload, and does that payload still identify a coalition member?

In this work, we systematically study multi-copy collusion attacks against neural audio watermarking systems. We evaluate five state-of-the-art systems and find that averaging only two valid personalized copies causes tracing failure in 87.7--99.7\% of trials while introducing little perceptual degradation. We further analyze two attribution outcomes after collusion: untargeted misattribution, where the decoded identity does not correspond to any contributing user, and targeted misattribution, where payload-aware weighting can steer the decoder toward a nonmember. Finally, we evaluate confidence-based screening as a potential defense.

Our contributions are summarized as follows. (1) We systematically evaluate multi-copy collusion across five neural audio watermarking systems and show that averaging only two valid copies can severely compromise traceability with limited audio degradation. (2) We analyze system-specific decoder behavior under untargeted and targeted misattribution. (3) We evaluate confidence screening as a mitigation strategy for unreliable attribution.

\begin{figure*}[!t]
  \centering
  \includegraphics[width=0.98\textwidth]{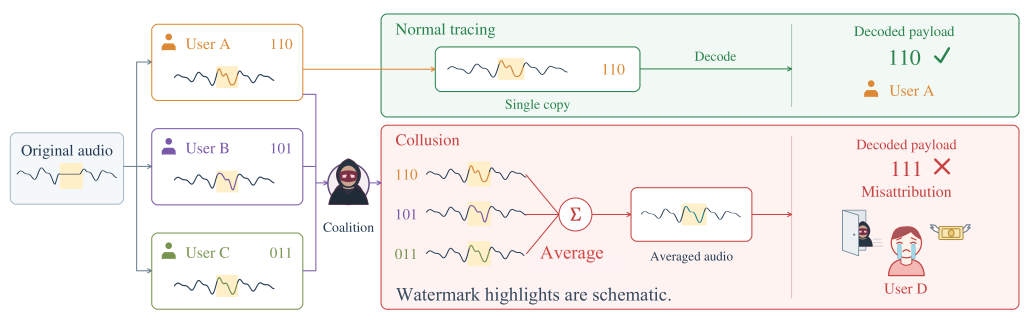}
  \caption{A single copy maps to its user; averaging personalized copies from a coalition can produce a payload assigned to a user outside the coalition.}
  \label{fig:scenario}
\end{figure*}

\section{Threat Model}
\label{sec:threat}

Each coalition member receives a personalized copy that carries a unique user-specific payload. We treat each payload in the system's native payload space as the identity of a distinct user. Our primary analysis uses aligned, uncompressed copies; Section~5.1 further evaluates temporal offsets and lossy coding before averaging. Under uniform averaging, the attack succeeds when the decoded payload matches none of the coalition payloads, leaving no colluding user traceable. The attacker uses only the coalition waveforms, without a clean reference, model parameters, or decoder queries.

Under targeted misattribution, we further consider a stronger payload-aware setting in which the attacker aims to make the decoded payload equal a nonmember payload selected in advance. In addition to the watermarked waveforms, the attacker knows the coalition payloads and candidate target payloads, then uses them to choose mixture weights. We define tracing outcomes at the payload level. Tracing succeeds if the decoder returns a coalition member's payload and fails otherwise. A failure is classified as \emph{targeted misattribution} when the output matches the prespecified nonmember target, and as \emph{untargeted misattribution} when it matches a different nonmember. We report outputs in each system's native payload space, and their eventual mapping to user identities depends on registry coverage, rejection policy, and the treatment of unregistered payloads.

We also explore a defense designed to detect and reject collusion outputs while accepting valid single-copy outputs. The defender has access to the native decoder and its confidence scores and can accept or reject an output before assigning the decoded payload to a user.

\section{Collusion Attacks}
\label{sec:attacks}

\subsection{Uniform Averaging}
\label{sec:averaging_attack}

Let $x$ be the shared speech waveform, $m_i$ the payload assigned to user $i$, and $w(x,m_i)=y_i-x$ the resulting watermark residual. User $i$ receives $y_i=x+w(x,m_i)$. For a coalition $\mathcal C$ of $K$ members, uniform averaging gives
\begin{equation}
\bar y_K=\frac{1}{K}\sum_{i\in\mathcal C}y_i
=x+\frac{1}{K}\sum_{i\in\mathcal C}w(x,m_i).
\label{eq:mean}
\end{equation}
The shared waveform remains in Eq.~\eqref{eq:mean}, while averaging combines the payload-dependent watermark residuals.

\subsection{Target-Bit Margin}
\label{sec:bit_margin}

\emph{Target-Bit Margin} uses only the coalition and candidate target payloads to optimize mixture weights toward a selected nonmember, without querying the decoder. Let $b_{ij}\in\{0,1\}$ be bit $j$ of coalition member $i$, $t_j$ the target bit, and $a_i$ the weight on that member's copy. The signed support for target bit $j$ is
\begin{equation}
\gamma_j=(2t_j-1)\left(\sum_i a_i b_{ij}-0.5\right).
\label{eq:margin}
\end{equation}

Target-Bit Margin maximizes a soft minimum of the target-bit margins with an entropy regularizer over convex mixture weights. To prevent solutions dominated by only a few copies, we require the effective coalition size $K_{\mathrm{eff}}=(\sum_i a_i^2)^{-1}$ to satisfy $K_{\mathrm{eff}}\geq\rho K$. For each target, the optimized weights produce the mixture $y=\sum_i a_i y_i$.

\section{Experimental Setup}
\label{sec:evaluation}

\subsection{Systems and Data}

We evaluate three bit-based systems---AudioSeal and WavMark with 16-bit payloads and TimbreWM with 10-bit payloads---and two latent-based systems, VoiceMark and WMCodec, both with 16-bit payloads \cite{sanroman2024audioseal,chen2023wavmark,liu2024timbre,li2025voicemark,zhou2025wmcodec}. Bit-based systems provide bit scores directly, whereas latent-based systems recover latent classes before mapping them to payload bits. We evaluate 300 ten-second recordings from 100 speakers, with three recordings per speaker and equal representation from AISHELL-3 and LibriSpeech \cite{shi2021aishell3,panayotov2015librispeech}.

\begin{table*}[!t]
\centering
\caption{Tracing-failure (TF) rate and audio quality under uniform averaging (300 trials per cell; for all 3 metrics, higher is better).}
\label{tab:risk}
\setlength{\tabcolsep}{3.7pt}
\renewcommand{\arraystretch}{0.94}
\begin{tabular}{l*{12}{c}}
\toprule
& \multicolumn{3}{c}{$K=2$}
& \multicolumn{3}{c}{$K=3$}
& \multicolumn{3}{c}{$K=5$}
& \multicolumn{3}{c}{$K=8$} \\
\cmidrule(lr){2-4}\cmidrule(lr){5-7}\cmidrule(lr){8-10}\cmidrule(lr){11-13}
System
& TF  (\%) & PESQ  & STOI 
& TF  (\%) & PESQ  & STOI 
& TF  (\%) & PESQ  & STOI 
& TF  (\%) & PESQ  & STOI  \\
\midrule
AudioSeal
& 99.0 & 4.60 & 0.999
& 97.3 & 4.59 & 0.999
& 98.3 & 4.59 & 0.999
& 100.0 & 4.58 & 0.999 \\
WavMark
& 96.7 & 4.57 & 0.999
& 97.7 & 4.56 & 0.999
& 98.0 & 4.54 & 0.998
& 99.3 & 4.52 & 0.998 \\
TimbreWM (10-bit)
& 87.7 & 4.56 & 0.998
& 84.7 & 4.56 & 0.997
& 88.0 & 4.55 & 0.997
& 92.0 & 4.53 & 0.996 \\
VoiceMark
& 93.7 & 4.20 & 0.983
& 99.0 & 4.17 & 0.978
& 99.3 & 4.11 & 0.974
& 98.7 & 4.05 & 0.971 \\
WMCodec
& 99.7 & 4.37 & 0.989
& 97.7 & 4.30 & 0.986
& 99.7 & 4.26 & 0.984
& 99.7 & 4.24 & 0.983 \\
\midrule
\multicolumn{13}{l}{\textit{Ideal bit-choice reference}} \\
16-bit
& 97.9 & \multicolumn{2}{c}{--}
& 99.5 & \multicolumn{2}{c}{--}
& 99.9 & \multicolumn{2}{c}{--}
& 99.9 & \multicolumn{2}{c}{--} \\
10-bit
& 89.1 & \multicolumn{2}{c}{--}
& 94.3 & \multicolumn{2}{c}{--}
& 96.8 & \multicolumn{2}{c}{--}
& 97.4 & \multicolumn{2}{c}{--} \\
\bottomrule
\end{tabular}

\end{table*}

Embedding, averaging, and decoding use each system's native sampling rate: 16~kHz for AudioSeal, WavMark, and VoiceMark; 22.05~kHz for TimbreWM; and 24~kHz for WMCodec.

\subsection{Evaluation Protocol and Metrics}

Coalition sizes are $K\in\{2,3,5,8\}$. For every recording, system, and $K$, we use distinct payloads to generate one personalized copy per member, and every copy is decoded correctly before mixing. At $K=5$, we shift one coalition copy at a time by $\pm10$, $\pm20$, or $\pm50$~ms before averaging; separately, we encode the copies using MP3 at 128~kbps or Opus at 64~kbps before averaging. The tracing-failure rate is the proportion of trials that result in tracing failure, whereas the target success rate is the proportion of attempts whose decoded payload matches the selected nonmember. PESQ and STOI measure mixture similarity to the first valid personalized copy in each coalition \cite{rix2001pesq,taal2011stoi}.

For Target-Bit Margin, we use an inverse temperature of 8, an entropy weight of 0.05, and $\rho=0.6$, fixed across systems and coalition sizes. We rank all nonmember payloads by their optimized soft-minimum margins and select the ten highest-ranked targets for each trial. For each $K\in\{5,8\}$, the target success rate is the fraction of the $300\times10$ attempts whose decoded payload exactly matches the selected target. Full optimization details are provided in the GitHub repository.

\section{Results and Analysis}
\label{sec:main_results}

\subsection{Tracing Failure under Uniform Averaging}

We first evaluate tracing failure under uniform averaging. Table~\ref{tab:risk} shows high tracing-failure rates across all five systems, often close to the ideal reference. The most important case is $K=2$: only two valid copies are sufficient to cause tracing failure in 87.7--99.7\% of trials, while PESQ and STOI indicate limited audio degradation.

This high TF is not an artifact of ideal alignment. At $K=5$, the mean TF remains above 96.5\% across all tested offsets. Quality scores decrease with larger offsets, but this decline does not necessarily indicate an equivalent loss of perceptual quality. Under MP3 and Opus coding, the mean TF remains between 96.5\% and 96.9\%, with PESQ above 4.25 and STOI above 0.989. Detailed results are available in the GitHub repository. We next examine why uniform averaging produces such high tracing-failure rates.

\subsection{Response to Coalition Bit Support}
\label{sec:composition}

We analyze \emph{bit support}, defined as the number of coalition payloads carrying 1 at a given position. In the \emph{ideal bit-choice reference}, each output bit is selected uniformly at random from coalition members, following the classical interleaving rule~\cite{laarhoven2014capacities}. The probability that an output bit is 1 is $c/K$, where $c$ is the bit support. Each bit comes from a coalition member, but the resulting payload may belong to none of them. We compare decoded outputs with this reference to identify where they differ from the interleaving rule.

\begin{figure}[!htbp]
\centering
\includegraphics[width=\columnwidth]{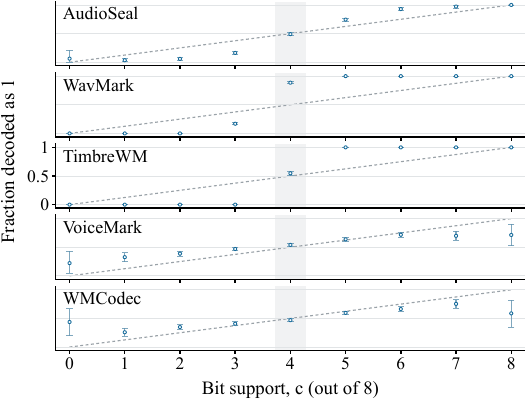}
\caption{Fraction decoded as 1 versus bit support at $K=8$. The dashed line is the ideal bit-choice reference; error bars show 95\% intervals across speakers.}
\label{fig:composition}
\end{figure}

Figure~\ref{fig:composition} shows that the five neural decoders do not all follow the linear $c/K$ relation predicted by the interleaving reference. The decoded-one rate rises with bit support in AudioSeal and WavMark, while TimbreWM follows the coalition majority. In VoiceMark and WMCodec, it changes less and remains closer to 0.5.

``Bit agreement'' measures agreement with the coalition majority at non-tied positions; ``Trial agreement'' measures the percentage of trials where all non-tied positions follow the majority. Table~\ref{tab:k8} shows higher agreement in bit-based systems than in latent-based systems. TimbreWM follows the majority at every non-tied position in all trials, yet its tracing-failure rate reaches 92.0\% at $K=8$ (Table~\ref{tab:risk}). Thus, even consistent majority decoding can produce nonmember payloads.

\begin{table}[!t]
\centering
\caption{Per-bit and full-trial majority agreement at $K=8$.}
\label{tab:k8}
\setlength{\tabcolsep}{5.0pt}
\begin{tabular}{ccc}
\toprule
System & Bit agreement $\uparrow$ (\%) & Trial agreement $\uparrow$ (\%) \\
\midrule
AudioSeal & 84.9 & 20.7 \\
WavMark   & 95.0 & 54.3 \\
TimbreWM  & 100.0 & 100.0 \\
VoiceMark & 61.9 & 0.7 \\
WMCodec   & 62.5 & 1.0 \\
\bottomrule
\end{tabular}

\end{table}

\subsection{Cross-Bit Coupling}
\label{sec:bit_coupling}

We examine whether mixing affects only the bit at which two payloads differ or also affects bits on which they agree. We compare AudioSeal, a bit-based system, with VoiceMark, a latent-based system. For each system, we use two payloads, A and B, that differ at only one position. We vary the mixing weight from 0 to 1 for 300 pairs. ``Other'' means the decoded payload matches neither A nor B.

\begin{figure}[!t]
\centering
\includegraphics[width=\columnwidth]{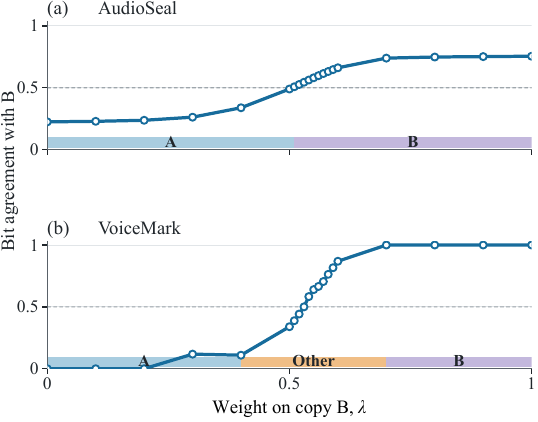}
\caption{Representative one-bit mixture paths. Curves show decoder support for B's value at the differing bit; color strips show the complete decoded payload (A, B, or Other).}
\label{fig:paths}
\end{figure}

In 299 of 300 trials, AudioSeal returns only A or B while VoiceMark frequently returns neither payload when the mixing weight is close to 0.5. Because A and B agree at every other position, these outputs necessarily contain changes in shared bits.

\subsection{Targeted Misattribution}
\label{sec:tampering}

The three bit-based systems respond predictably to coalition bit support. Using Target-Bit Margin, we next test whether an attacker in the stronger payload-aware setting can control the decoded identity rather than merely evade tracing.

\begin{table}[!htbp]
\centering
\caption{Target success rate and audio quality.}
\label{tab:targets}
\begin{tabular}{ccccc}
\toprule
& \multicolumn{2}{c}{Target success rate $\uparrow$ (\%)}
& \multicolumn{2}{c}{Average quality} \\
\cmidrule(lr){2-3}
\cmidrule(lr){4-5}
System & $K=5$ & $K=8$ & PESQ $\uparrow$ & STOI $\uparrow$ \\
\midrule
AudioSeal & 41.9 & 43.6 & 4.58 & 0.999 \\
WavMark   & 51.4 & 54.4 & 4.52 & 0.998 \\
TimbreWM  & 78.4 & 92.3 & 4.52 & 0.996 \\
VoiceMark &  1.1 &  0.6 & 4.04 & 0.972 \\
WMCodec   &  1.0 &  1.3 & 4.20 & 0.983 \\
\bottomrule
\end{tabular}

\end{table}

Table~\ref{tab:targets} shows high target success rates for AudioSeal, WavMark, and TimbreWM, with TimbreWM reaching 92.3\% at $K=8$. VoiceMark and WMCodec remain near 1\%, despite their high tracing-failure rates under uniform averaging. This contrast mirrors their responses in Figure~\ref{fig:composition}: Target-Bit Margin is most effective when decoded bits vary predictably with coalition support. Target-Bit Margin exploits bit-level support; whether latent-aware attacks can achieve stronger control remains open.

\section{Rejecting Unreliable Assignments}
\label{sec:confidence}

As a preliminary mitigation, we examine whether native decoder confidence can be used to reject unreliable collusion outputs.

\emph{Bit confidence} is the decoder's support for the selected bit value. We use native bit probabilities where available, WavMark's valid-window vote fraction, and the summed probabilities of latent classes consistent with each bit for VoiceMark and WMCodec. We summarize each output by its minimum confidence, mean confidence, and the logarithm of the variance of confidence across bits.

\begin{figure}[!htbp]
\centering
\includegraphics[width=\columnwidth]{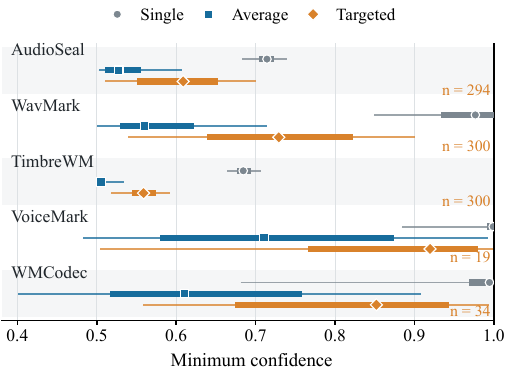}
\caption{Minimum confidence for Single, Average, and successful Targeted outputs at $K=8$. For Targeted, \(n\) denotes the number of trials contributing one successful target hit.}
\label{fig:confidence}
\end{figure}

Figure~\ref{fig:confidence} shows the distributions of minimum confidence for valid single-copy outputs (\emph{Single}), uniformly averaged mixtures (\emph{Average}), and successful targeted mixtures (\emph{Targeted}) at $K=8$.

We evaluate the screening rule using five-fold cross-validation. In each fold, 80\% of the speakers are used for calibration and 20\% for testing. An output is accepted only when its mean and minimum confidence values are above their respective lower thresholds and its log variance is below its upper threshold. The mean-confidence and log-variance thresholds are estimated using Gaussian fits, whereas the minimum-confidence threshold is set using an empirical quantile. The thresholds are calibrated so that the resulting screening rule retains at least 95\% of the valid single-copy outputs on the calibration speakers.

After screening, a \emph{Targeted} attempt counts as successful only if the decoder attributes its output to the selected nonmember and the output passes the screen. The denominator used to calculate the target success rate remains unchanged after screening.

\begin{table}[!htbp]
\centering
\caption{Confidence screening at $K=8$.}
\label{tab:rejection}
\begin{tabular}{cccc}
\toprule
System
& \shortstack{Single\\accepted $\uparrow$ (\%)}
& \shortstack{Average\\rejected $\uparrow$ (\%)}
& \shortstack{Target success\\rate $\downarrow$ (\%)} \\
\midrule
AudioSeal & 94.7 & 100.0 & 43.6 $\rightarrow$ 2.4 \\
WavMark   & 94.7 & 100.0 & 54.4 $\rightarrow$ 4.4 \\
TimbreWM  & 95.0 & 100.0 & 92.3 $\rightarrow$ 0.0 \\
VoiceMark & 94.3 & 77.7  & 0.6 $\rightarrow$ 0.3 \\
WMCodec   & 95.0 & 88.0  & 1.3 $\rightarrow$ 0.7 \\
\bottomrule
\end{tabular}

\end{table}

The screen accepts about 95\% of Single outputs and rejects all Average outputs from the three bit-based systems. Most Average outputs from the latent-based systems are also rejected. After screening, target success rates fall to at most 4.4\%.

\section{Conclusion}

Across five systems, averaging only two personalized copies severely degrades traceability. The failure persists under temporal offsets and lossy coding. We further evaluate targeted misattribution attacks, which can cause the three bit-based systems to attribute collusion outputs to selected nonmembers. This contrast distinguishes attribution escape from control over the resulting identity. We also explore a preliminary defense using three decoder-confidence statistics to reject most collusion outputs. These findings motivate including multi-copy collusion in neural audio watermarking evaluations.

\bibliographystyle{IEEEbib}
\bibliography{references}

\end{document}